\documentstyle[preprint,aps]{revtex}
\input psfig.sty

\def\ltsim{\vbox {\hbox{\lower 1.4\baselineskip \hbox{$<$}} \break
		 \hbox{\lower 0.2\baselineskip \hbox{$\sim$}} } }
\def\k{{\bf k}}
\def\p{{\bf p}}
\def\pp{{\bf p}^\prime}
\def\ppp{{\bf p}^{\prime\prime}}
\def\kp{{\bf k}^\prime}

\def\q{{\bf q}}

\def\R{{\bf R}}

\def\wt{\tilde \omega}
\def\wtp{\tilde\omega^\prime}
\def\wtpp{\tilde\omega^{\prime\prime}}
\begin{document}
\draft


\title{
Scattering by impurity-induced order parameter ``holes'' in d-wave
superconductors }

\author{
Matthias H. Hettler $^{a,b}$ and P.J. Hirschfeld$^a$ 
}

\address{
$^a$Dept. of Physics, Univ. of Florida, Gainesville, FL 32611, 
USA.\\
$^b$Dept. of Physics, Univ. of Cincinnati, Cincinnati, OH 45221-0011
}

\maketitle
\begin{abstract}
Nonmagnetic impurities in d-wave superconductors cause strong local 
suppressions of the order parameter.  We 
investigate the observable effects
of the scatterigng off such suppressions
 in bulk samples by treating the order parameter 
``hole" as a pointlike off-diagonal scatterer treated within a self-consistent 
t-matrix approximation.   Strong scattering potentials lead to a finite-energy 
spectral feature in the d-wave ``impurity band", the observable effects of
which include enhanced low-temperature microwave power absorption and a 
stronger sensitivity of the London penetration depth to disorder than 
found previously in simpler ``dirty'' d-wave models.
\end{abstract}
{\it Introduction.}  The current consensus regarding the d-wave
symmetry of the order parameter in the hole-doped cuprates rests to
a considerable extent on the 
success of simple models of the effect of disorder on unconventional
superconductors.\cite{Reviews}     Theories based on the impurity
t-matrix approximation\cite{HVW,SMV,joynt} suggest that disorder 
gives rise to an ``impurity band'' of extended states near the Fermi level
which dominates the low-temperature behavior of
the system.  Although published data on  superconducting
properties continue to exhibit a certain variability, such
apparent discrepancies can frequently
be accounted for within this picture by allowing for the  variation
in sample quality.   Systematic substitution of Copper 
with nonmagnetic planar defects such as $Zn$ or $Ni$ leads to similar 
effects.  The observed strong temperature  dependence of transport 
properties of  nominally
pure samples with nearly optimal
$T_c$  gives rise to the further conclusion of near-unitarity
limit scattering by most simple defects in these systems.  

Several puzzles and discrepancies must be 
understood before the problem of disorder in the
superconducting state of the high-temperature 
superconductors (HTSC's) may be considered solved
even at a purely phenomenological level.  The 
simplest theory of low-temperature transport in a
d-wave superconductor\cite{HPS} predicts an $\omega\rightarrow 0$ 
conductivity $\sigma\simeq \sigma_{00}
+aT^2$, where $\sigma_{00}\simeq ne^2/(m\pi\Delta_0)$ is a 
``universal'' constant dependent on the scale set by the order
parameter maximum $\Delta_0$ but independent of the relaxation 
rate $1/\tau$ for weak disorder.\cite{PALee}  Experiments on 
untwinned samples clean enough to exhibit the linear-$T$
magnetic penetration depth characteristic of a pure d-wave state down
to a few degrees are consistent with the universal residual conductivity
value,\cite{Zhangetal} but exhibit a  conductivity
temperature variation $\delta\sigma\simeq T$ or even sublinear
temperature dependence.\cite{BonnHardy}

Further difficulties include the rather slow measured rate
at which the critical temperature itself is supressed by 
impurities, (for planar defects, roughly a factor of two slower 
than predicted by the simplest theory) as well as a considerably more rapid
decrease of the low-$T$ superfluid density $\rho_{s}$ with
disorder than predicted.\cite{Zhangetal,Carbotte} 
The two effects together result in a
ratio $T_c/\rho_s$ which is considerably larger than predicted by
``dirty d-wave" theory in many HTSC samples, as pointed out
recently by Franz et al.\cite{Franzetal}  These authors argued that
the usual theory, which assumes a spatially homogeneous
disorder-averaged order parameter (OP) $\Delta_\k$,  neglects an important
physical effect, namely the supression of the true order
parameter $\Delta_\k({\bf R})$ around the positions of the defects.
These static fluctuations are expected to be large since the order
parameter is d-wave in character, and highly local since the order
parameter varies (crudely speaking) on a scale of the coherence length
$\xi$, which at low $T$ is of order a few \AA \, in HTSC's.  In a full
numerical solution to the Bogoliubov-de Gennes (BdG) equations for a d-wave
system with dilute strong potential scatterers, Franz et al. indeed
observed an enhancement of $T_c/\rho_s$ consistent with experiment.
Very recently\cite{zhit-walk}, a more analytic approach to the effect
of order parameter suppression  by Zhitomirsky and Walker 
found a reduction of the $T_c$-suppression in rough agreement 
with experimental observations. Both approaches work best close
to $T_c$. On the other hand, the discrepancy with  experiment in the 
$T$--dependence of the transport properties is most evident at 
low temperatures. Clearly, a complementary approach for low $T$ 
is needed.

In the interest of constructing a practical theory of transport in
the HTSC's including these order parameter suppressions, we have
attempted to exploit the short range nature of the order parameter supression
(at low $T$) 
by replacing the true, self-consistently determined $\Delta_\k({\bf R})$
near the impurity site by a pointlike order parameter ``hole" which acts as
an off-diagonal (in the Nambu--Gorkov matrix sense)
potential scatterer for electrons.  The weight of
the off-diagonal perturbation is then determined by the bare impurity
potential and by the solution to the one-impurity problem for the
${\bf R}$-integrated OP suppression.  This approximation makes sense at 
low temperatures, mainly because the zero $T$ coherence length which
controls the range of the OP suppressions 
is at most a few atomic lengths for the HTSC's.
With this approximation,
one can perform disorder averages in the usual way and obtain a
generalization of the very flexible t-matrix approximation involving a
translationally invariant, effective medium.

Having outlined the
method, we proceed to show that the failures of the
``dirty d-wave" model when compared to experiments on  low-$T$ microwave
conductivity and the dependence of the absolute penetration depth 
on disorder are at least partially cured.   The results for the
{\it temperature} dependence of $\lambda$ and other thermodynamic quantities
does change very little. 
A preliminary account of this research  has appeared in Ref. 
\cite{hettlerthesis}.

{\it Local order parameter supression by single impurity.} Several authors
have solved the BdG equations or equivalent for the local structure of the
d-wave
$\Delta_\k({\bf R})$ around a nonmagnetic impurity.\cite{singleimp} 
In our approach, we make the usual BCS assumption that the  pairing potential 
is separable, $V_{k,k'}\equiv V \Phi(k) \Phi(k')$, 
and has d-symmetry,  with $\Phi_d(\k)\equiv \sqrt{2} \cos 2\phi$ normalized 
over a model circular Fermi surface.
We further neglect so-called ``leading loser" components of the pair
interaction, e.g. subdominant pairing channels which have been shown to lead
to additional fine structure in the order parameter around the impurity site
despite being energetically forbidden in the bulk. 

The bare impurity itself is described for simplicity by a $\delta$-function
potential in real space, $\hat U({\bf R}-{\bf R}_{imp})=U_0\delta({\bf R}-
{\bf R}_{imp})\tau_3$, where the $\tau_i$ are the Pauli matrices in
particle-hole space.  Initially, one might try to find the spatial
variation of the order parameter, which is given by the BCS gap equation after
subtraction of the bulk limit.  The Fourier transform of the (static) order
parameter fluctuation
$\delta\Delta_\k(\q)$ is then determined\cite{Rusinov,hettlerthesis} by the
single-impurity t-matrix $\hat{T}(\p,\pp)$,
\begin{eqnarray}
&&\delta\Delta_\k(\q)=  V \Phi_d(\k) T \sum_{\omega} \sum_{{\bf k'}}
\Phi_d({\bf k'}) \times\\
&&\mbox{Tr}\left\{\frac{\tau_1}{2} \hat{G_0}({\kp+\q}/2)
\hat{T}(\kp+\q/2,\kp-\q/2) \hat{G_0}({\k'-q/2})\right\}.\nonumber\\
\label{ddelzero}
\end{eqnarray}
where $\hat G_0$ is the matrix Green function of the pure system. 
The t-matrix is given as usual by
\begin{equation}
\hat{T}(\p,\pp)= \hat{U}(\p,\pp) +\sum_{\ppp} \hat{U}(\p,\ppp)
\hat{G}_0(\ppp) \hat{T}(\ppp,\pp),
\label{tmatdef}
\end{equation}
where $\hat{T}(\p,\pp)$ and $ \hat{U}(\p,\pp) $ are Fourier transforms with
respect to the electronic momenta. 
In the usual ``dirty d-wave" theory,  the t-matrix is taken independent of
momentum,
$\hat T = \hat  T(\omega)$, for the case of isotropic scatterers.   Here,
we explicitly account for the fact that electrons moving in the neighborhood of
the impurity feel an effective one-body potential due not only to the bare
impurity but to the order parameter modification about the impurity.
The effective impurity potential is therefore
\begin{equation}
\hat{U}_\k(\q)=\hat U_0\tau_3 +\delta\Delta_\k(\q)\tau_1
\label{potential}
\end{equation}
where $\k=(\p+\pp)/2$ is the  momentum conjugate to the ``fast'' relative
motion of the electron pair, while $\q=\p-\pp$ is the momentum
conjugate to the slowly varying center of mass position.  Solving 
Eqs. \ref{ddelzero}--\ref{potential}
self-consistently is equivalent to solving the BdG equations for a single
impurity, and will yield a momentum-dependent t-matrix in the
d-wave case.  The associated order parameter fluctuation $\delta\Delta_\k(\q)$ 
 has d-wave symmetry
in real space about the impurity site.\cite{singleimp}

Our objective instead is to develop a formalism for calculating 
observables in the
presence of a finite concentration of impurities.  To this end, we neglect
the $\q$-dependence of $\delta\Delta_\k(\q)$ in Eq. \ref{potential}
equivalent to
replacing the full $\delta\Delta_\k(\R-\R_{imp})$ with
$\delta\Delta_\k(\q=0)\,\,\delta(\R-\R_{imp})$.  This is certainly a
reasonable  approximation in the cuprates at temperatures sufficiently far
below $T_c$, since the very short $T=0$ coherence length $\xi_0$ on which the
order parameter varies becomes comparable to the lattice constant.  
We then solve Eq. \ref{ddelzero} for $\delta\Delta_\k(\q=0)$ under the 
continued assumption of negligible subdominant pair component, such that
$\delta\Delta_\k(\q=0)=\delta_d\cos 2\phi$. We point out that 
no new parameter has been introduced into the theory by this procedure, since
$\delta_d$ is driven by the impurity scattering strength $U_0$.

\begin{figure}[h]
\leavevmode\centering\psfig{file=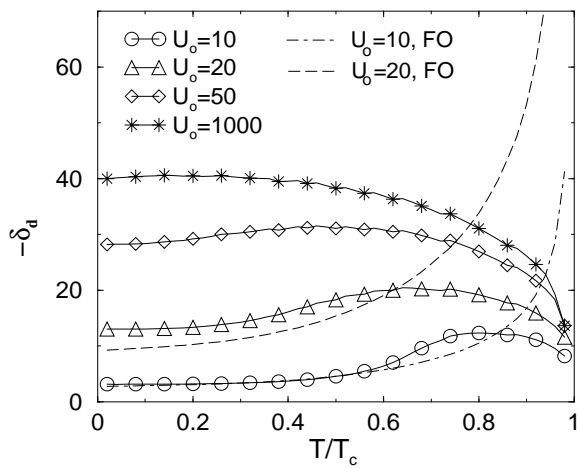,width=3. in}
\caption{Order parameter scattering strength vs. temperature.
The dot-dashed and dashed lines are the first order (FO) result 
\protect \cite{ChoiMuzikar} for 
$\delta_d$ with  $U_0=10$ and $U=20$, respectively, the
symbols and solid lines are the
self-consistent results for $U_0$=10,20,50,1000,  as indicated.
For this example,
$(N_0)^{-1}=100 T_c$ and $\Delta_o=3 T_c$. 
The pairing potential and the cutoff are chosen such that the
BCS--expression for $T_c$ is the unit of energy, 
$T_c= 1.14 \omega_o \exp{(-1/N_o|V|)} \,
\,\, (\omega_o=30,|V|\sim 28.31$)}
\label{ddelfig}
\end{figure}
We begin by solving Eqs. \ref{ddelzero}--\ref{potential} numerically under 
the approximations outlined above. In Fig. 
\ref{ddelfig} we show the temperature dependence of
the self-consistent evaluation of $\delta_d$ in comparison
 with the result obtained when the $\tau_1$ part of the
t-matrix is iterated only once.\cite{ChoiMuzikar}  Close to $T_c$, 
the first order (FO) result  diverges like $(1-T/T_c)^{-1/2}$ 
whereas the self-consistent
result vanishes  with the bulk order parameter $\Delta_0$ like 
$(1-T/T_c)^{1/2}$. For low  $T$ and the smallest $U_0=10$ the first order
and the self-consistent result agree well, as $\delta_d$ is sufficiently 
small. But already for $U_0=20$ the discrepancies are obvious 
even for zero temperature. For small $U_0$ (and low $T$) the first order result 
underestimates $|\delta_d|$, however, for large $U_0$ it overestimates  $|\delta_d|$ (not shown in the figure). 
Note that $\delta_d$ can be 
quite large in the unitarity limit $U_0\rightarrow \infty$. 
However, it can never be infinite as its $U_0$--dependence saturates like 
$U_0^2/(1+const U_0^2)$. In a similar fashion the nonlinear corrections 
in $\delta_d$ prevent the divergence of  $\delta_d$ 
at the critical temperature (see the expression for the t-matrix below).

{\it Self-energies in t-matrix  approximation.}  
The solution to Eq. \ref{tmatdef} 
at $\q=0$ in the present ansatz may be written
\begin{eqnarray}
\hat T_\k(\omega)={U_0^2g_0+U_0\tau_3\over 1-U_0^2g_0^2}+
\delta_d{\delta_dg_0+(1-\delta_dg_2)\cos 2\phi\tau_1\over
(1-\delta_dg_2)^2-(\delta_dg_0)^2},
\label{tmatres}
\end{eqnarray}
where $g_0$ and $g_2$ are the components of the momentum integrated Green 
function, $g_0\equiv (1/2)\sum_k{\rm Tr} {\hat G}(\k,\omega)$ and
$g_2\equiv (1/2)\sum_k{\rm Tr}\, \tau_1 \cos 2\phi\, \hat{ G}(\k,\omega)$.
This solution is not exact but is obtained under the reasonable 
assumption of the smallness
of higher order scattering terms which have no inhomogeneous driving term.
\cite{hettlerthesis}
The disorder-averaged self-energy is now defined in the limit of independent
scattering centers to be $\hat \Sigma(\k,\omega)\equiv n_i \hat T_\k (\omega)$,
and determined self-consistently with the averaged $\hat G$ via the Dyson
equation, 
$\hat G^{-1}=\omega-\xi_k\tau_3-\Delta_k\tau_1-\hat\Sigma(\k,\omega)\equiv
\wt-\tilde \xi_k\tau_3-\tilde\Delta_\k\tau_1$. 
The first term in Eq. \ref{tmatres} 
is the result obtained in the usual dirty d-wave theory
for arbitrary scattering phase shift
$\delta_0=-\cot^{-1}(1/N_0U_0)$.  For strong scattering,
$\delta_0\simeq \pi/2$, a resonance occurs at or near the Fermi level, 
as can be seen by examining the
corresponding denominator; this leads to the finite density of states at the
Fermi level and ``gapless" behavior, as has been discussed extensively in the
literature.\cite{HVW,SMV}

The denominator in the second term, due to off-diagonal
scattering, leads to a similar resonance.  To estimate the
position  of this feature, we examine the clean limit, in which $g_2\simeq
-\pi N_0[2/\pi +i(\omega^2/\Delta_0^2)\log (4\Delta_0/\omega)]$ and $g_0\simeq
-\pi N_0[\omega\log (4\Delta_0/\omega)+i\omega/\Delta_0]$ for
$\omega/\Delta_0\ll 1$.  The resonance in the second term of Eq. \ref{tmatres}
then occurs 
when $\omega/\Delta_0 =[-2/\pi+(\pi\delta_d N_0)^{-1}]\equiv \tilde c_f$.   Our
self-consistent determination of $\delta_d$ given above shows that as 
$U_0\to \infty$, $\tilde c_f \simeq 0.2$, so the resonance from the  
off-diagonal channel never occurs at the Fermi level.  It is easy to check that
on resonance the imaginary part of the denominator is of the same order
as the real part, up to log corrections.  We therefore expect that the
resonance will modify the low-energy behavior of the quasiparticle relaxation
time at the lowest energies in the clean limit.

\begin{figure}[h]
\leavevmode\centering\psfig{file=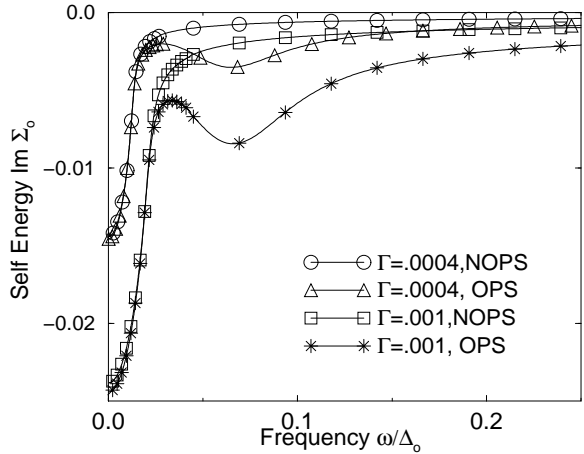,width=3. in}
\caption{${\rm Im} \Sigma_0(\omega)$ for several values of $\Gamma$, with and
without OP scattering. 
The resonance at finite energy is a direct consequence of the 
additional OP scattering.
The zero frequency resonance is qualitatively unchanged.}
\label{sig03}
\end{figure}

These notions are confirmed by the full numerical evaluation of 
the self-energies. In Fig. \ref{sig03} we show the imaginary part of 
the diagonal self-energy for
various values of the impurity scattering rate parameter 
$\Gamma\equiv n_i/\pi N_0$.  In the clean limit the two resonances are clearly
distinguishable, but they merge with increasing disorder. 
This is because the zero frequency feature
grows only  $\propto\gamma\sim\sqrt{\Gamma\Delta_o}$ but the feature
at finite frequencies $\propto\Gamma$ (at least initially).  Thus, above a
certain disorder we no longer expect off-diagonal scattering to 
qualitatively modify low-temperature transport.  There should, however,
be novel temperature-dependent effects at small disorder.  In addition,
the disorder dependence of any quantity which is sensitive 
to all energy scales, as, e.g. the
$T=0$ superfluid density discussed below, may be substantially modified.

{\it Microwave conductivity and penetration depth.}
In the limit of vanishing external frequency $\Omega=0$, 
the quasiparticle conductivity is given by\cite{HPS}
\begin{equation}
\sigma(T,\Omega=0) = -\frac{n e^2}{ m } \int_{-\infty}^{\infty}
d\omega \left(-\partial f\over \partial \omega\right ) S(\omega,T),
\label{sigma1}
\end{equation}
where $n$ is the electron density, $e$ and $m$ the electron charge and mass,
respectively, and 
\begin{eqnarray}
S(\omega,T)=\int {d\phi\over 4\pi}\left[ {\rm Im}{\wt^2\over
\xi_+^3} - {{1}\over {2}} \frac{{\tilde\omega}^\prime_+{}^2 +{\tilde\Delta}_{\k+}^{\prime\prime}{}^2} {{\wtpp}_+ {\wtp}_+ -{\tilde\Delta}_{\k+}^{\prime\prime} {\tilde\Delta}_{\k+}^\prime}
\,{\xi_+^\prime\over \,|\xi_+|^2}\right],
\label{sigma2}
\end{eqnarray}
where $\xi_\pm\equiv \pm \sqrt{\wt_\pm^2-\tilde\Delta_{\k\pm}^2}$, the
subscripts $\pm$ indicate evaluation at $\omega\pm i0^+$, and real and 
imaginary parts are denoted by $^\prime$ and $^{\prime\prime}$, respectively. 
The first term in Eq. \ref{sigma2} gives
rise  to the ``universal" $T\to 0$ conductivity $\sigma_{00}$,
while the second term determines the low-temperature behavior. 
\begin{figure}[h]
\leavevmode\centering\psfig{file=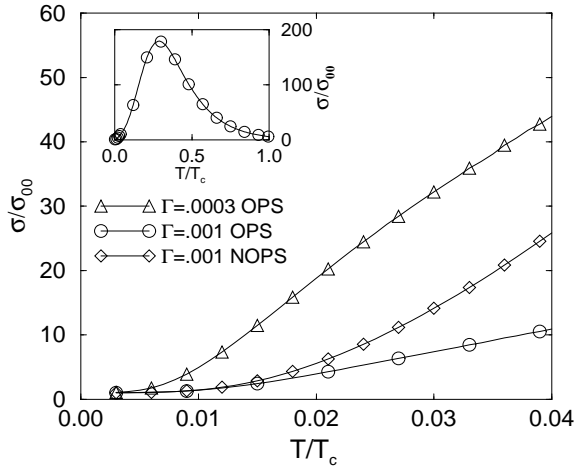,width=3. in}
\caption{Low-$T$ microwave conductivity at various disorder  $\Gamma$, 
with full $T$-range in inset. Note the quasi--linear behavior of the 
data with OP scattering. The conductivity without OP scattering is
always quadratic in $T$ at low temperatures.}
\label{mrek3}
\end{figure}
In Figure \ref{mrek3}, we show the results of a numerical evaluation of 
\ref{sigma1} with and without  OP scattering.
In the absence of OP scattering,
$\sigma\simeq  T^2$ for almost the entire region (the conductivity is measured
in units of the "universal" zero temperature value $\sigma_{00}$). The values 
of $\Gamma$ and $\Delta_o=3T_c$ have been
chosen by fitting the low temperature penetration depth of recent experiments 
on YBCO  single crystals.\cite{BonnHardy,HPS}. 
This result for the conductivity
is at odds with the observed dominant linear behavior observed in the same 
samples.
In contrast, the result including OP scattering  displays quasi-linear 
behavior above an energy scale that is
a fraction of $\gamma$ due to the new resonance. This becomes more obvious 
if we lower the parameter $\Gamma$ in the case of additional OPS, so that the 
conductivity at higher temperatures is in rough agreement with the standard 
theory and the experimental data (not shown in the figure). 
The additional feature in the self energy 
can increase the conductivity by more than a factor of two at the appropriate
temperatures, and therefore resembles  much more the dominant linear behavior
of the experimental data.

In the inset to Fig. \ref{mrek3}, we show the full temperature range of the 
conductivity using the model of Ref. \cite{HPS} for the inelastic scattering.
The peak in $\sigma$ and subsequent drop at higher T is due to the competition
between  impurity and inelastic scattering.

\begin{figure}[h]
\leavevmode\centering\psfig{file=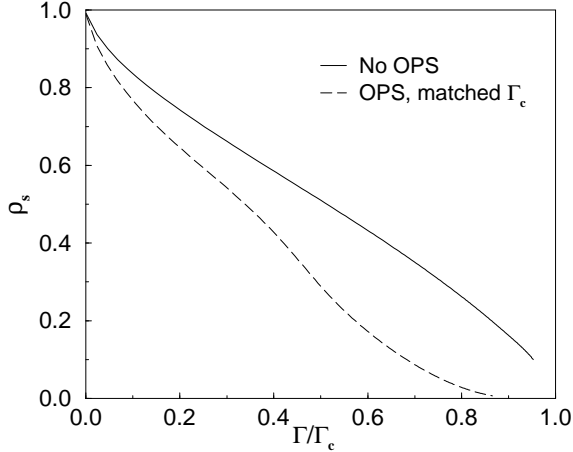,width=3. in}
\caption{ Superfluid density $\rho_s$ vs. $\Gamma$ with \& w/o OP scattering.
In order to eliminate trivial effects due to the suppressed critical 
concentration in the OP scattering case we scale the corresponding data
so that the $\Gamma_c$ are the same $\Gamma_c \sim 0.82$. 
Note the increased initial
slope in  the OP scattering case. Close to $\Gamma_c$ our
approximate treatment of  OP scattering becomes invalid.
}
\label{rhovsgam}
\end{figure}

The superfluid density and penetration depth are given by evaluating 
the imaginary part of the conductivity at $\omega=0,q\to 0$ if we ignore
nonlocal corrections which can 
\vskip -.3cm
\noindent be important for $T\ltsim 1K$ in very clean
samples:\cite{KosztinLeggett}
\begin{equation}
\rho_s=-\int_0^\infty d\omega \tanh {\omega\over 2T}\int {d\phi\over 2\pi} {\rm
Re} {\tilde
\Delta_\k^2\over (\wt^2-\tilde\Delta_\k^2)^{3/2}}
\end{equation}
Because we have accounted for a new source of scattering, we expect a faster
depletion of the superfluid density for a given impurity concentration.  In
Figure \ref{rhovsgam}, we show the dependence  of $\rho_s$ at $T=0$ on 
disorder with and without OP scattering. 
The comparison is complicated by the fact that two cases have
different critical concentrations ( the concentration at which 
$T_c$ is suppressed to zero temperature). The $T_c$ (and consequently
the critical concentration) including
OP scattering is much suppressed. This is  an artifact of our
treatment of the  OP scattering which is invalid when the range
of the OP suppression becomes large (close to $T_c$ or the critical 
concentration).
In order to make the comparison meaningful we scale the result of the
OP scattering case  so that $\Gamma_c = n_c/\pi N_o$ is the same as in the
standard case without OP scattering.

 The increase in the initial slope of the
$\rho_s$ suppression over the usual dirty d-wave approach
is about 30--40\%.  This is less than  the difference 
noted in Ref. \cite{Carbotte}
in damaged YBCO films.  We are not aware of 
measurements of the absolute scale of $\rho_s$ in which the samples are
systematically disordered by, e.g. $Zn$ substitution for $Cu$.
Such experiments would be very useful to confirm the importance of order
parameter scattering for bulk properties.

Low temperature properties dependent primarily on the density of states 
(DOS) should be largely unaffected by the new source of scattering. 
For example,
the $T\to T^2$ crossover with disorder in the penetration depth of a d-wave
superconductor
\cite{felds} is controlled by the  finite residual density of states at
zero energy, $N(\omega\to 0)$, that arises due to impurities.  As this
residual DOS is
determined primarily by the resonance  of the self energy
at the Fermi level in the 
unitarity limit, little change from the standard theory is observed.
{\it Conclusions.}
The strong
suppression of the order parameter around nonmagnetic impurity sites is
an important qualitative difference between d-wave systems and classic 
s-wave superconductors, where such suppressions are small.
The spatial structure of the quasiparticle states and condensate
in the neighborhood of an impurity in the Cu-O plane will soon be probed with
STM as has been achieved for the similar vortex problem.  We have posed
the complementary question of how disorder in the off-diagonal channel
induced by impurities modifies bulk average properties of a d-wave
superconductor.  By making the plausible assumption of {\it local}
off-diagonal scattering, we have constructed a simple theory able to
describe these effects.  We have shown that the measured quasi-linear
temperature dependence of the microwave conductivity at low temperatures
can be obtained within our theory, as a result of a finite-energy
scattering resonance which arises in the unitarity limit.  We have
also shown that the suppression of the $T=0$ superfluid density
with disorder is
enhanced over the usual theory, although the temperature
dependence  is only weakly affected.  Systematic measurements of the
disorder dependence of $\lambda (T\to 0)$ or $\rho_s(T \to 0)$ will aid in
establishing the validity of our picture.


\indent
{\it Acknowledgments.}  The authors gratefully acknowledge 
extensive
discussions with A.S. Balatsky, M. Franz, and D.J. Scalapino and 
A. Zhitomirsky.  
Partial support was provided by
NSF grants DMR-96--00105, DMR-97--04021 and DMR-93--57199.

\end{document}